\documentclass[a4paper, amsfonts, amssymb, amsmath, reprint, showkeys, nofootinbib, twoside]{revtex4-1}
\usepackage[english]{babel}
\usepackage[utf8]{inputenc}
\usepackage{textgreek}
\usepackage[colorinlistoftodos, color=green40, prependcaption]{todonotes}
\usepackage{amsthm}
\usepackage{xcolor}
\usepackage{graphicx}
\usepackage{cases}
\usepackage[T1]{fontenc}

\usepackage[pdftex, pdftitle={Article}, pdfauthor={Author}]{hyperref} 
\usepackage[normalem]{ulem}
\usepackage{scalerel}
\usepackage{tikz}
\usetikzlibrary{svg.path}

\usepackage{xcolor}

\definecolor{orcidlogocol}{HTML}{A6CE39}
\tikzset{
  orcidlogo/.pic={
    \fill[orcidlogocol] svg{M256,128c0,70.7-57.3,128-128,128C57.3,256,0,198.7,0,128C0,57.3,57.3,0,128,0C198.7,0,256,57.3,256,128z};
    \fill[white] svg{M86.3,186.2H70.9V79.1h15.4v48.4V186.2z}
                 svg{M108.9,79.1h41.6c39.6,0,57,28.3,57,53.6c0,27.5-21.5,53.6-56.8,53.6h-41.8V79.1z M124.3,172.4h24.5c34.9,0,42.9-26.5,42.9-39.7c0-21.5-13.7-39.7-43.7-39.7h-23.7V172.4z}
                 svg{M88.7,56.8c0,5.5-4.5,10.1-10.1,10.1c-5.6,0-10.1-4.6-10.1-10.1c0-5.6,4.5-10.1,10.1-10.1C84.2,46.7,88.7,51.3,88.7,56.8z};
  }
}

\newcommand\orcidicon[1]{\href{https://orcid.org/#1}{\mbox{\scalerel*{
\begin{tikzpicture}[yscale=-1,transform shape]
\pic{orcidlogo};
\end{tikzpicture}
}{|}}}}

\begin{document}
\title{Unimodular Gravity with Arbitrary Diffusion Function: \\
A Dynamical System Reconstruction Approach}

\author{Gabriel G\'omez$^1$\orcidicon{0000-0002-3618-9824}}
\email{luis.gomezd@umayor.cl}

\author{Guillermo Palma$^2$\orcidicon{0000-0001-7326-964X}}
\thanks{corresponding author: guillermo.palma@usach.cl}

\author{Norman Cruz$^{2,3}$\orcidicon{0000-0002-0737-3497}}
\email{norman.cruz@usach.cl}

\affiliation{$^1$ Centro Multidisciplinario de F\'isica, Vicerrector\'ia de Investigaci\'on, Universidad Mayor, \\ Camino La Pir\'amide 5750,  Huechuraba, 8580745, Santiago, Chile.\\
$^2$Departamento de F\'{\i}sica, Universidad de Santiago de Chile, Avenida Ecuador 3493, Santiago, Chile,\\ 
$^3$Center for Inter-disciplinary Research in Astrophysics and Space Exploration (CIRAS), Universidad de Santiago de Chile, Av. Libertador Bernardo O’Higgins 3363, Estaci\'on Central, Chile.}

\date{\today} 

\begin{abstract}

We investigate cosmological diffusion models in unimodular gravity within a dynamical systems reconstruction framework. By treating the logarithmic slope of the diffusion sector as an invertible dynamical variable, the diffusion function can be systematically reconstructed from the phase-space structure of the cosmological evolution. Under these conditions, we determine the physically admissible fixed points of the system, identifying novel matter--diffusion scaling solutions associated with power-law diffusion sectors, as well as purely diffusion-dominated configurations capable of driving late-time accelerated expansion without requiring an explicit cosmological constant term. The local behavior around the fixed points is then extended to the full cosmological evolution, providing a framework to explore the global implications of diffusion cosmologies. Beyond the asymptotic fixed-point structure, we further develop a reconstruction formalism based on the dynamical evolution of the diffusion slope, allowing for trajectories interpolating between different diffusion regimes during the cosmic history. Our results establish a systematic framework for constructing and classifying viable diffusion cosmologies in unimodular gravity directly from the phase-space dynamics.

\end{abstract}


\maketitle

\section{Introduction}
\label{sec:intro}

The persistent theoretical and observational challenges of modern cosmology have motivated the exploration of alternative gravitational frameworks, among which unimodular gravity (UG) has received renewed attention \cite{jirouvsek2023unimodular,alvarez2024origin,smolin2009quantization, kaplan2025redshifting, Plaza:2025hde, Gielen:2024bho}. From a theoretical standpoint, the cosmological constant problem remains one of the most profound open puzzles in physics \cite{Weinberg:1988cp}. Within the UG framework, the restricted diffeomorphism invariance of the Einstein–Hilbert action — which preserves the volume element \cite{Einstein:1919gv,Anderson:1971pn,Buchmuller:1988wx,Unruh:1988in,Henneaux:1989zc,Ng:1990xz,Finkelstein:2000pg,Ellis:2010uc} leads to Einstein's field equations in which the cosmological constant arises naturally as an integration constant. This problem has also been tackled in modified theories of gravity \cite{kibarouglu2024cosmology, Salvio:2024ugq,Nojiri_2016, Odintsov_2016}.

A special feature of this framework is the occurrence of a modified conservation law, which implies an energy diffusion process, characterized here by the energy diffusion function (EDF) $Q(t)$. 
One physically motivated choice for the EDF draws from the \textit{continuous spontaneous localization (CSL)} model~\cite{Pearle:1976ka,Ghirardi:1985mt,Pearle:1988uh,Ghirardi:1989cn}, in which energy is created through quantum collapse. The mass-proportional variant of CSL motivates an equation of state (EoS) that relates the EDF to the energy density of the matter fluid, naturally taking the form of a barotropic EoS
$Q(t):=\alpha_{i} \rho_{i}$, where  $\alpha_{i}$ is a constant and $\rho_{i}$ is the energy density of the different matter components ($i = b,\gamma,\nu,cdm,\Lambda$).  In this simple Ansatz, the diffusion is assumed to be proportional to the energy density of the fluid itself, and was first introduced in \cite{Corral:2020lxt} to investigate the late-time behavior of a FRW model containing dark matter and a dark energy term modeled by $\Lambda + Q(t)$, with $\Lambda>0$. The parameter $\alpha$, constrained to the case where only the dark matter fluid undergoes diffusion, was fitted to cosmological data, revealing that this model closely reproduces the behavior of $\Lambda$CDM model \cite{Corral:2020lxt}. Furthermore, since $Q(t)$ is a decreasing function, owing to the dilution of matter density as the universe expands, the model asymptotically approaches a de Sitter expansion. The implications of this Ansatz were further explored in the context of interacting fluids to alleviate the Hubble tension  \cite{perez2021resolving, LinaresCedeno:2020uxx}. 

More complex expressions for the EDF have been explored in \cite{garcia2019cosmic, garcia2021universe}, where $Q$ is a function of the jerk parameter, defined as the dimensionless third-order time derivative of the scale factor.

The energy diffusion process in UG gives rise to an effective cosmological constant (CC), which opens the possibility of an effective dynamical phantom behavior. This type of behavior has been favored by recent cosmological data~\cite{Riess_2018,2019NatAs...3..272R}. Thus, UG naturally produces novel phantom-like dynamics, motivating the investigation of their possible future singularities and associated characteristics. In Ref.~\cite{Cruz:2026tjt}, future cosmological singularities were investigated for diffusion models of the form $Q(a)=Q_{0}a^{\beta}$, which admit analytical solutions for the fluid energy density by solving the non-conservation equation inherent to the UG formalism. In addition, this class of diffusion functions leads to a positive production of cosmic entropy under suitable constraints on the model parameters \cite{cruz_exploring_2024}.

The EDFs discussed above represent different ways in which non-conservation can affect the evolution of the cosmic fluid and, consequently, the cosmological dynamics. Nevertheless, there is currently no widely accepted form for the EDF that can be derived from a well-established physical process. As a result, most proposed forms for $Q$ are motivated either by phenomenological considerations or by mathematical simplicity.

The dynamical systems approach is a powerful tool for understanding the global behavior of cosmological models \cite{bahamonde2018dynamical}, including those arising from UG. By introducing suitable phase-space variables, the cosmological equations can be recast as an autonomous system whose fixed points characterize the asymptotic regimes of the cosmic evolution. In this way, the phase-space structure can be used not only to determine the possible cosmological behaviors, but also to reconstruct admissible diffusion sectors directly from the underlying dynamics.

Motivated by this idea, the aim of the present work is to investigate, through dynamical systems techniques, how suitable diffusion functions $Q$ can be reconstructed directly from the cosmological phase-space structure, without assuming {\it a priori} a specific diffusion model. In particular, we focus on diffusion sectors capable of reproducing transitions between matter-dominated and accelerated expansion regimes. 

The paper is organized as follows. In Sect.~\ref{sec:diff}, we review the fundamental aspects of unimodular gravity relevant to the diffusion framework, together with the thermodynamic conditions required for positive entropy production. In Sect.~\ref{sec:dynamical_systems}, we derive the autonomous dynamical system associated with a general diffusion sector and establish the conditions required for its closure. We then investigate the cosmological fixed points and their physical interpretation in Sect.~\ref{sec:cosmology}, including their relation to existing diffusion scenarios. Building upon this phase-space structure, Sect.~\ref{sec:dif_rec} develops a general reconstruction framework for diffusion functions beyond asymptotic fixed-point regimes. Finally, we summarize our results and present our conclusions in Sect.~\ref{sec:conclusions}.

We set throughout this paper the units such that $c = 1 = \hbar$, and therefore $\kappa := \sqrt{8\pi G_{N}} =\sqrt{ 8\pi /M_{P}} $, where $G_{N}$ is the Gravitational constant, and $M_{P} = \sqrt{\hbar c / G_{N}}$ is the Planck mass.

\section{Background dynamics of Unimodular gravity}
\label{sec:diff}

In several formulations of UG, see, for instance, Refs.~\cite{Einstein:1919gv, BUCHMULLER1988292,PhysRevD.40.1048, 10.1063/1.529283,10.1063/1.1328077}, the gravitational dynamics is described by a restricted subset of the Einstein field equations. This subset, which comprises the trace-free Einstein equations, allows the full Einstein equations to be recovered, with the cosmological constant emerging as an integration constant. Within the UG framework, the dynamical equations governing the gravitational field are given by

\begin{equation}
R_{\mu \nu} - \frac{1}{4}g_{\mu \nu}R  = T_{\mu \nu}- \frac{1}{4}g_{\mu \nu}T, 
\label{eq:unieom}
\end{equation}
being $T$ the trace of the energy-momentum tensor of the matter fields, defined as usual.

A concrete realization of the field equations presented above is obtained by imposing the unimodular condition on the Einstein--Hilbert action through a Lagrange multiplier $\Lambda(x)$. 
 The Lagrangian density of the gravitational sector then takes the form $\mathcal{L} = \sqrt{-g}[R +\Lambda(x)(1-\zeta(x)/\sqrt{-g})]$, where $\zeta(x)$ is a scalar density. Varying the corresponding action --- constructed from the above Lagrangian density together with the matter fields --- with respect to the metric yields
\begin{equation}
    R_{\mu \nu} - \frac{1}{2}g_{\mu \nu}R + \Lambda(x) g_{\mu \nu}  = T_{\mu \nu}, \label{eq:unieom1}
\end{equation}
and the volume element is fixed through the condition $\zeta(x) = \sqrt{-g}$, which is obtained by performing the variation of the action with respect to $\Lambda(x)$. From the trace of equation (\ref{eq:unieom1}) one gets
\begin{equation}
    \Lambda(x) = \frac{1}{4}(R+T), 
    \label{eq:lagrange}
\end{equation}
which allows us to write (\ref{eq:unieom}) once we replace it back into (\ref{eq:unieom1}).
If we compute the divergence of equation (\ref{eq:unieom1}) with the Lagrange multiplier written in (\ref{eq:lagrange}), due to the Bianchi identity $\nabla^{\mu}G_{\mu \nu}=0$, being $G_{\mu \nu}:= R_{\mu \nu} - (1/2)g_{\mu \nu}R$ the Einstein tensor, we obtain the following
\begin{equation}
    \frac{1}{4}\nabla^{\mu}g_{\mu \nu}(R+T)=\nabla^{\mu}T_{\mu \nu}, \ \ \ \mbox{or} \ \ \ \nabla_{\nu}\Lambda(x) = \nabla^{\mu}T_{\mu \nu}.
\end{equation}
The above equation can be solved as
\begin{equation}
    \Lambda(x) = \Lambda_{0} + \kappa^2Q, \label{eq:diff}
\end{equation}
where where $\Lambda_{0}$ is an integration constant whose value is fixed by observations, and $Q(x)$
is an arbitrary function that quantifies the degree of violation of local energy-momentum conservation. For this reason, $Q(x)$ is commonly referred to as the {\it energy diffusion function} (EDF). It is worth noting that if the conservation of $T_{\mu \nu}$ is imposed as an additional condition, then $\Lambda(x)$  reduces to a standard cosmological constant \cite{calderon2021negative, akarsu2020graduated}. Otherwise, a non-constant $Q(x)$ gives rise to a dynamical effective cosmological term  $\Lambda(x)$. In the context of a spatially flat FLRW universe, and considering a barotropic cosmic fluid characterized by the equation of state\footnote{Since our focus is on the late-time cosmological evolution, the fluid will be identified with the dark matter sector. For simplicity, radiation is neglected throughout this work.} $p = w \rho$, where $w$ is the equation of state (EoS) parameter, the Friedmann equations in the UG framework take the form

\begin{align}\label{eq:fried1}
 3H^2 &= \kappa^2\left(\rho + Q \right) + \Lambda_{0} ,\\
 \label{eq:accel} 2\dot{H} + 3H^2 &= - \kappa^2\left(p-Q \right) + \Lambda_{0} ,
\end{align}
where an overdot denotes the derivative with respect to cosmic time and $H(t)=\dot{a}/a$ is the Hubble function. Since $\Lambda_{0}$  arises as an integration constant in the UG framework, all three cases $\Lambda_{0}<0,\Lambda_{0}=0$ and $ \Lambda_{0}>0$ are in principle admissible. Within the context of the observable universe, however, the value of $\Lambda$ has been observationally constrained using cosmological data in the UG framework \cite{Corral:2020lxt}. It is worth noting that a negative cosmological constant cannot be ruled out {\it a priori}, as it may play a relevant role in describing certain phases of cosmic evolution \cite{calderon2021negative, akarsu2020graduated, Visinelli_neg_CC_2019}. In this work, we restrict our analysis to the case $\Lambda_{0} > 0$
, in accordance with current observational evidence, leaving the exploration of the remaining cases for future work.

Combining Eqs.~(\ref{eq:fried1}) and (\ref{eq:accel}), one obtains the continuity equation for the energy density,
\begin{equation}
\dot{\rho } + 3H(\rho  + p )= - \dot{Q },\label{eq:nonconsrho}
\end{equation}
where the diffusion function $Q=Q(t)$ depends only on cosmic time in order to preserve the homogeneity and isotropy required by the cosmological principle.

Thus, once $Q(a)$ is defined, the matter density $\rho(a)$ can be obtained exactly from the continuity equation Eq.~(\ref{eq:nonconsrho}), thereby fixing the background cosmological dynamics. 

The continuity equation admits the exact solution
\begin{equation}
\rho(a) = \frac{\rho_0}{a^{3(1+w)}}
- \frac{1}{a^{3(1+w)}} \int_{a_0}^{a} \tilde{a}^{3(1+w)} \frac{dQ}{d\tilde{a}}\, d\tilde{a},\label{eq:sol_rho_matter}
\end{equation}
where $\rho_{0}=\rho(a_{0})$ denotes the present-day energy density. Hereafter we take $a_{0}=1$. This solution explicitly shows that deviations from the standard scaling $\rho \propto a^{-3}$ for the cold dark matter component arises from the cumulative energy exchange with the diffusion sector.

Finally, we discuss the conditions under which the diffusion function $Q$ produces a positive growth of the entropy ($dS/dt >0$). In particular, we follow the analysis presented in Ref. \cite{cruz_exploring_2024}. From the Gibbs equation given by \cite{Callen:450289}
\begin{equation}
    TdS = dU+pdV, \label{eq:gibss}
\end{equation}
where $U$ is the internal energy defined as $U:=\rho V$, $V = V_{0}a^{3}$ is the volume and $T$ the temperature of the cosmic fluid. We obtain in terms of the diffusion function the following expression \cite{cruz_exploring_2024}
\begin{equation}
 T\frac{dS}{da} =-V\frac{dQ}{da}. \label{eq:entropy}
\end{equation}
Notice that, in the case $Q=\mbox{constant}$, the cosmic expansion evolves adiabatically, in agreement with the standard cosmological model. Therefore, the second law of thermodynamics is satisfied provided that
\begin{equation}
    \frac{dQ}{da} \leq 0,
    \qquad \Rightarrow \qquad
    \frac{dS}{da} \geq 0,
\label{eq:good}
\end{equation}
which, equivalently, implies $\dot{Q}\leq0$. This condition is completely general and does not depend on the particular form of the diffusion function. In particular, it does not restrict the sign of $Q(t)$ itself. Through Eq.~(\ref{eq:diff}), the diffusion sector may contribute either positively or negatively to the dynamical cosmological term, thereby affecting the cosmological expansion in qualitatively different ways.

\section{Dynamical System Formulation}
\label{sec:dynamical_systems}
In order to analyze the cosmological dynamics we introduce the dimensionless variables
\begin{equation}
\Omega\equiv\frac{\kappa^{2}\rho}{3H^2}, 
\qquad
\Omega_Q\equiv\frac{\kappa^{2} Q}{3H^2},
\qquad
\Omega_{\Lambda_{0}}\equiv\frac{\Lambda_{0}}{3H^2}.\label{eqn:dimensionless_variables}
\end{equation}
These variables satisfy the Friedmann constraint
\begin{equation}
\Omega+\Omega_Q+\Omega_{\Lambda_{0}}=1 .\label{eqn:Friedmann_constr}
\end{equation}
A more intuitive and common choice would be to treat the diffusion function and the cosmological constant as a single effective contribution, parametrized as $\Omega_{\Lambda} \equiv \Lambda(t)/(3H^{2})$, where $\Lambda(t) \equiv \Lambda_{0} + \kappa^{2} Q(t)$. However, keeping these two components separate, as in Eq.~(\ref{eqn:dimensionless_variables}), allows us to isolate the genuine dynamical impact of diffusion throughout the cosmic evolution\footnote{Since the diffusion function enters only the continuity equation for $\rho$ and does not directly source the evolution of $\Lambda$, the structure of the autonomous system remains unchanged.}. This distinction is particularly useful, as the evolution of the diffusion sector directly influences the dynamical properties of the system.

We will use the e-fold number $N= ln(a)$ as an independent time-like variable, so that $'$ denotes differentiation with respect to $N$.

The acceleration equation in terms of the phase-space variables becomes
\begin{equation}
\frac{H'}{H}=-\frac{3}{2}(1+w)\Omega .
\label{H_prime_over_H}
\end{equation}

To characterize the dynamical evolution, we introduce the effective equation-of-state parameter
\begin{equation}
w_{\rm eff} = -\left(1+\frac{2H'}{3H}\right) = (1+w)\Omega - 1.
\end{equation}
This expression makes the physical behavior transparent: accelerated expansion, $w_{\rm eff} < -1/3$, can be achieved whenever the energy budget is dominated by the diffusion component, the constant $\Lambda$, or a combination of both.

In order to characterize the expansion properties associated with the different cosmological fixed points, it is also useful to introduce the deceleration parameter. Expressed in terms of the phase-space variables, it takes the form
\begin{equation}
q := - \frac{\dot{H}}{H^2} - 1 = \frac{3(1+\omega)}{2}\Omega - 1 .
\label{decel_par}
\end{equation}
%
\subsection{Autonomous system}

Using the non-homogeneous continuity equation and the definitions above, the cosmological dynamics can be written in terms of the phase-space variables $(\Omega, \Omega_{Q}, \lambda)$ as the autonomous system
\begin{align}
\Omega' &= \lambda(N)\,\Omega_Q +3(1+w)\Omega(\Omega-1), \label{eqn:Omega_evol}\\
\Omega_Q' &= -\Omega_Q\left[\lambda(N)-3(1+w)\Omega\right], \label{Omega_Q_evol}\\
\lambda' &= F(\lambda),\label{eqn:evol_slope_diffusion}
\end{align}
where we introduced the diffusion slope variable $\lambda$, and the function $F(\lambda)$ by
\begin{equation}
\lambda(N) := -\frac{Q^{\prime}}{Q}, \quad \text{and} \quad F(\lambda) = \lambda^2\left( 1 - \Gamma_{Q} \right), 
\label{F_lambda}
\end{equation}
and $\Gamma_{Q}$, which measures the curvature of the diffusion trajectory in logarithmic space, is

\begin{equation}
\Gamma_Q \equiv \frac{Q Q''}{Q'^2}.\label{eqn:def_curvature}
\end{equation}
In the above evolution equations we have assumed that $F$ can be expressed as a function of $\lambda$, allowing the cosmological equations to be cast into autonomous form. This requirement imposes a restriction on the admissible diffusion functions $Q$, since the slope parameter $\lambda$ must constitute a valid dynamical variable. In particular, $\lambda$ must be expressible as a single-valued function along the dynamical trajectories, which implies that $\lambda(N)$ must be locally invertible. By the inverse function theorem, this holds in any interval where $d\lambda / dN \neq 0$. Hence, if $d\lambda / dN $ does not change sign in the domain of interest, then $\lambda(N)$ is monotonic and therefore globally invertible on that domain. 

Together with the Friedmann constraint, Eq.~(19), the system evolves in a reduced phase space spanned by $(\Omega,\Omega_Q,\lambda)$. The hypersurfaces $\Omega_Q=0$ and $\lambda=0$ define invariant manifolds of the dynamics. Moreover, the evolution equation for the diffusion slope $\lambda$ decouples from the remaining phase-space variables, revealing a hierarchical structure in the dynamical system. In particular, one can first determine the evolution of $\lambda(N)$ independently, and subsequently solve the remaining reduced cosmological system on the subspace $(\Omega,\Omega_Q)$.

Further, for $\Gamma_Q=1$, the evolution equation for $\lambda$ becomes identically satisfied, so that the diffusion slope remains constant. The resulting reduced dynamical system corresponds to a two-dimensional autonomous system.

A first example of admissible diffusion functions satisfying the invertibility condition is obtained by considering a constant diffusion slope, $\lambda=\beta$, with $\beta$ constant. In this case, Eq.~(\ref{eqn:def_curvature}) can be integrated straightforwardly, yielding the power-law diffusion function
\begin{equation}
Q(a)=Q_0 a^{-\beta}.
\end{equation}
This class of diffusion models has previously been introduced from phenomenological considerations \cite{Cruz:2026tjt}; here, however, it emerges naturally from the phase-space structure of the dynamical system.

More general monotonic choices for the diffusion slope are also possible. For instance, taking $\lambda = \lambda_{0} / (1 + \beta N)$ leads to the diffusion function
\begin{equation}
Q(N)= \frac{Q_{0}}{(1+\beta N)^{\lambda_{0}/\beta}}.
\end{equation}
Such generalized diffusion sectors allow for a dynamical evolution beyond the simple power-law behavior and will be further analyzed in the diffusion reconstruction framework developed in Sect.~\ref{sec:dif_rec}.

\subsection{Fixed Points and their stability properties}

The fixed points correspond to the stationary points of the autonomous system defined by equations (\ref{eqn:Omega_evol}-(\ref{eqn:evol_slope_diffusion}). Moreover, since the r.h.s. of Eq.~(\ref{eqn:evol_slope_diffusion}) vanishes identically for a power-law diffusion function, the stability analysis of the fixed points can be performed starting from the reduced dynamical system (\ref{eqn:Omega_evol}-(\ref{Omega_Q_evol}), together with the condition $\lambda = \beta$. 
To this aim, we first compute the derivatives of the functions on the r.h.s. of the dynamical system to obtain the Jacobian matrix. Then, from its characteristic eigenvalues we extract the stability properties of the fixed points. Accordingly, we obtain the following expressions:
\begin{align}
F_{1,\Omega}&= 3(1+\omega)(2\Omega-1), \quad F_{1,\Omega_Q}= \beta, \label{F1_der}\\
F_{2,\Omega}&= 3(1+\omega)\Omega_Q, \quad F_{2,\Omega_Q}= - \left[ \beta - 3(1+ \omega)\Omega \right] .\label{F2_der}
\end{align}
\subsubsection{Matter dominated solution}

A first fixed point is obtained for
\begin{equation}
\Omega = 1, \quad \Omega_{Q}=0, \quad \text{ and } \quad \Omega_{\Lambda_{0}} = 0.
\end{equation}
Since diffusion effects are negligible, in this regime the cosmological dynamics reduces to the standard matter-dominated expansion without acceleration, which follows from Eq.~(\ref{decel_par}), or $q = (1 + 3\omega) / 2 > 0$. From Eqs.~(\ref{F1_der}-\ref{F2_der}) we conclude from the eigenvalues of the Jacobian matrix
\begin{equation}
    \mu_1= 3(1+\omega), \quad \mu_2 = -\beta + 3(1+\omega),
\end{equation}
that this point does not correspond to an attractor, but rather to a saddle point. \\
Moreover, from Eq.(\ref{eq:accel}), the Hubble parameter behaves in a neighborhood of this fixed point as
\begin{equation}
H \propto a^{-\frac{3~(1 + \omega)}{2}}.
\end{equation}
In the vicinity of the fixed point and for $\omega \approx 0$, this expression reproduces the standard cosmological evolution of cold dark matter.

\subsubsection{De Sitter solution}

A second interesting fixed point occurs for
\begin{equation}
\Omega_{\ast}=0, \qquad \Omega_{Q} = 0, \quad \text{and} \quad \Omega_{\Lambda_{0}} = 1 .
\end{equation}

In this case the expansion becomes asymptotically de Sitter with $H^2 \approx \Lambda_{0}/3$ and, therefore $\omega_{eff} = -1$. The universe is entirely dominated by the cosmological constant, while both matter and diffusion contributions vanish asymptotically. Now, from Eqs.~(\ref{F1_der}-\ref{F2_der}), we obtain the eigenvalues of the Jacobian matrix
\begin{equation}
    \mu_1= -3(1+\omega), \quad \mu_2 = -\beta,
\end{equation}
which shows that this point does correspond to an attractor for $\beta > 0$.

\subsubsection{Scaling matter-diffusion solution}

Considering the alternative branch $\Omega_{Q}\neq0$, a nontrivial fixed point arises when 
\begin{equation}
\lambda = 3(1 + \omega) ~\Omega_{\ast},
\label{lambda_vs_Omega}
\end{equation}
together with
\begin{equation}
\Omega_{\ast} + \Omega_Q = 1.
\label{scaling_sol}
\end{equation}

From Friedmann's equation, this implies $\Omega_{\Lambda_{0}} = 0$, so the energy budget of the universe is shared between matter and the diffusion sector. As for the above fixed points, the eigenvalues of the Jacobian matrix are obtained from Eqs.~(\ref{F1_der}-\ref{F2_der}), leading to
\begin{equation}
    \mu_{1} =  \beta -3(1+\omega), \quad \text{and} \quad \mu_{2} = \beta.
\end{equation}
The above eigenvalues show that this point behaves as a repeller for $\beta > 3(1+\omega)$, and as a saddle point for $0 < \beta < 3(1+\omega)$. \\

In this regime, $Q$ evolves proportional to the DM energy density, i.e. $Q \approx r \rho$, where $r \in (0,1)$ is a constant ratio. Now, from the continuity equation for the DM energy density, Eq.~(\ref{eq:nonconsrho}), we conclude that 
\begin{equation}
\rho \propto a^{-3(1 + \omega) /(1+r)}. 
\label{rho_asymp}
\end{equation}
The above expression corresponds to the regime in which the diffusion slope remains constant. In this case, one has
\begin{equation}
\lambda = - \frac{Q^{\prime}}{Q} \approx - \frac{\rho^{\prime}}{\rho} \approx \frac{3(1 + \omega)}{1+r},
\end{equation}
showing that the diffusion function and the matter density evolve with comparable scaling behavior along the cosmological trajectory. Hence, the non-dynamical character of $\lambda$ guarantees that the stationarity condition $\lambda^{\prime}=0$ is satisfied.

Moreover, from the expression of the deceleration parameter $q$ given in Eq.~(\ref{decel_par} ), we infer that the condition  %
\begin{equation}
\frac{(1 + 3~\omega)}{2} < r < 1,
\end{equation}
corresponds to an accelerated expansion of the universe. Remarkably, this behavior arises solely from the diffusion term, without the need for a standard cosmological constant contribution. However, this solution describe a transient stage of the cosmological evolution preceding the asymptotic late-time state.

Finally, from the above expression for the dark matter energy density $\rho$ and the Raychaudhuri equation, we infer that the asymptotic behavior of the Hubble parameter is
\begin{equation}
H \propto a^{-\frac{3}{2}(1+w)\Omega_{\ast}},
\end{equation}
which makes explicit that the expansion rate is controlled by the constant matter fraction $\Omega_{\ast}$. This scaling implies that the diffusion component tracks the matter density throughout the cosmological evolution.

As a conclusion of this case, we emphasize that the occurrence of accelerated expansion within this scaling regime indicates that diffusion effects can remain dynamically relevant without becoming the dominant component, thereby providing a mechanism to sustain a constant fractional dark matter contribution throughout the cosmological evolution.

\subsubsection{Diffusion dominated solution}

The previous fixed point corresponds to a regime in which diffusion effects are dynamically significant, although they do not completely dominate the total energy budget. In contrast, one may further consider the pure diffusion regime, characterized by
\begin{equation}
\Omega = 0, \qquad \Omega_Q = 1, \qquad \text{and} \qquad \Omega_{\Lambda_{0}} = 0.
\end{equation}
Similarly to the previous fixed points, the eigenvalues of the Jacobian matrix are obtained from Eqs.~(\ref{F1_der}--\ref{F2_der}), yielding
\begin{equation}
    \mu_1=-\left[\beta+3(1+\omega)\right], \qquad
    \mu_2=0.
\end{equation}
The first eigenvalue is negative for $\beta<3(1+\omega)$, indicating that the fixed point is locally attractive along the corresponding direction. However, the second eigenvalue vanishes, implying that the critical point is non-hyperbolic. Consequently, linear stability theory alone cannot determine its asymptotic behavior along the remaining direction. To fully characterize its stability, we complemented the linear analysis with a numerical study of the phase-space trajectories. The resulting flow shows that, although trajectories are attracted toward the critical point along one direction, they are repelled along the nonlinear direction. See Figs.~\ref{fig:phase_space} and \ref{fig:evol_cosmo}. Therefore, the diffusion-dominated solution is not a late-time attractor but rather a saddle point of the dynamical system.

In order to ensure that this configuration represents a stationary point 
in phase space, we must impose the condition
\[
\lambda(N_{*}) = 0,
\]
for some value $N_{*}$ characterizing the fixed point.

From the definition of the diffusion slope parameter $\lambda(N)$,
the condition $\lambda=0$ implies that the diffusion function becomes constant,
\begin{equation}
Q = Q_0 .
\end{equation}
Using the Friedmann equation, one obtains that the Hubble parameter becomes constant
\begin{equation}
H_{Q} = \sqrt{\frac{Q_0}{3}} .
\end{equation}
This solution corresponds to a de Sitter-like phase in which the scale factor undergoes exponential expansion,
\begin{equation}
a(t) \propto e^{H_{Q}t}.
\end{equation}
Physically, a constant diffusion function behaves effectively as a cosmological constant with $\omega_{\mathrm{eff}}=-1$. In this regime the diffusion sector dominates the energy budget and dynamically drives accelerated expansion without requiring a fundamental vacuum energy component. Diffusion therefore acts as an effective dark energy source at late times.

In summary, the matter-dominated and de Sitter solutions exist independently of the specific form of the diffusion function, since they lie on invariant submanifolds where the diffusion sector does not contribute to the dynamics. 

More generally, whenever $\lambda=0$ or $F(\lambda_\ast)=0$—with $\lambda_\ast\neq0$ denoting a nontrivial root of $F(\lambda)$—the corresponding fixed points arise independently of the explicit form of the diffusion function, although their existence and stability remain tied to the global phase-space structure. As discussed in Sect.~\ref{sec:dif_rec}, the curvature function $\Gamma_Q(\lambda)$ plays a central role in determining the dynamical interpolation between these asymptotic cosmological regimes.

\section{From fixed points to physically suited diffusion-function Ansätze} \label{sec:cosmology}

In the previous section, we have identified the fixed points that reproduce the standard cosmological evolution. In the vicinity of these phase-space configurations, the diffusion function is not arbitrary. Instead, it is constrained by the dynamical system, which determines its local functional behavior.

In this section, we investigate the physical consequences of extending the analytically determined behavior of the diffusion function, derived near the fixed points, to the full cosmological evolution. This strategy is motivated by the fact that the fixed points of the autonomous dynamical system encode the relevant cosmological epochs. Consequently, any viable diffusion function must be consistent with the existence of these fixed points and reproduce their associated local behavior in the corresponding asymptotic regimes.

\subsubsection{Power-law diffusion Ansatz associated with the matter-diffusion scaling solution} 

Since the scaling matter–diffusion solution implies that, near the fixed point, the diffusion function scales proportionally to the matter density (see discussion above Eq.~(\ref{rho_asymp})), we promote this local behavior to a global Ansatz within the UG framework considered here:

\begin{equation}
Q(a) = \alpha \; \rho(a)
\label{Q_propto_rho}
\end{equation}
with $\alpha$ being a dimensionless constant. Integrating the continuity equation for the matter density one obtains
\begin{equation}
\rho(a) = \rho_{0} \; a^{-\beta}    
\end{equation}
where $\beta = 3(1 + \omega) / (1 + \alpha)$. Inserting this expression into the Ansatz for $Q$, we conclude that the diffusion function follows a power law behavior, i.e. $Q(a) = \alpha \; \rho_{0} \; a^{-\beta}$. Using the thermodynamic requirement of positive entropy production, one obtains the condition
\begin{equation}
 \alpha > 0 \qquad \text{or} \qquad \alpha < -1 .
\end{equation}
However, consistency with the standard cosmological evolution further restricts the allowed parameter space. Since the matter density must decrease as the universe expands, only positive values of $\alpha$ lead to a physically viable cosmological behavior.

The time evolution of the Hubble parameter is obtained directly from the Raychaudhuri equation, or equivalently from Friedmann equation as:
\begin{equation}
H^2(a) = \frac{\kappa^2}{3} \Bigl[\; \rho_{\Lambda_{0}} + (1+ \alpha) \; \rho_{0} \; a^{-\beta} \;\Bigr],    
\end{equation}
where $\rho_{\Lambda_{0}}= \Lambda_{0} / \kappa^2$. The model parameters can be expressed in terms of the present-day Hubble parameter through
\begin{equation}
3 H_{0}^2 = \kappa^2 \rho_{0}\,(1+\alpha) + \Lambda_{0} ,
\end{equation}
which provides an observational constraint on the admissible values of the diffusion parameter $\alpha$. This scenario was confronted with observations in Ref.~\cite{Corral:2020lxt}, yielding the constraint $\alpha>0$. Physically, a positive value of $\alpha$ indicates that the diffusion process induces a gradual transfer of energy away from the dark matter sector as $Q(t)$ evolves.

Since the dark matter density decreases as the universe expands, the effective cosmological contribution,
$\rho_{\Lambda}(a)\equiv \rho_{\Lambda_{0}}+Q(a)$, 
also evolves with cosmic time, gradually approaching the constant background value $\rho_{\Lambda_{0}}$ as the diffusion term becomes subdominant at late times. In this way, the effective cosmological contribution tracks the matter evolution during part of the cosmological history, alleviating the coincidence problem within the UG framework for the Ansatz under consideration. 

Since $H^2$ must remain non-negative, the above expression imposes a constraint on the allowed values of $\Lambda_{0}$, which may be either positive or negative. In particular, for $\Lambda_{0}>0$ the asymptotic cosmological evolution is compatible with a de Sitter phase
\begin{equation}
 a(t) \xrightarrow{t\to\infty} \exp(-H_{\infty}t)     
\end{equation}
with 
\begin{equation}
 H_{\infty} = \left( \; \Lambda_{0}/3 \;\right )^{1/2}.   
\end{equation}
Nevertheless, when $\Lambda_{0} < 0$ the above expression implies the existence of an upper bound on the scale factor,
\begin{equation}
    a(t) \leq a_{upper} = \Bigl[\frac{-\Lambda_{0}}{\kappa^2 (1+\alpha) \rho_{0}}\Bigr]^{(1+\alpha) / 3\gamma},
\end{equation}
beyond which the Hubble parameter ceases to be well defined. This solution corresponds to a universe whose cosmological future terminates in a big crunch.

Finally, integrating the defining relation for the scale factor $a$ in terms of the Hubble parameter $H(a)$,
\begin{equation}
t-t_{0} = \int^{a}_{a_{0}} \frac{ da / a}{ H(a)} 
\label{scale_parameter}
\end{equation}
where $a_{0} = a(t_{0})$, and 

\begin{equation}
H(a)= \left[H_{0}^2 + \frac{\kappa^2 \; (1+ \alpha) \; \rho_{0}}{3} \; (a^{-\beta} -1)\right]^{1/2},
\label{Hubble_a}
\end{equation}
the scale factor is implicitly determined as a function of cosmic time $t$. For special values of the parameters $\alpha$ and $\beta$, the integral admits an analytic evaluation via asymptotic expansions, and the resulting cosmological scenarios were analyzed in detail in Ref.~\cite{cruz_exploring_2024}.

\subsubsection{Constant Ansatz for the diffusion function}

Following the same reasoning adopted for the scaling matter--diffusion Ansatz, we now investigate the cosmological consequences of extending the constant diffusion solution to the entire cosmological evolution:
\begin{equation}
Q = Q_{0}.
\end{equation}
From a physical perspective, a constant diffusion function can be interpreted as an effective contribution to the cosmological-constant sector. In this case, one may define an effective cosmological constant as
\begin{equation}
\widetilde{\Lambda} = \Lambda_{0} + \kappa^2 Q_{0}.
\end{equation}
Accordingly, the model reduces to the standard cosmological scenario with an effective cosmological constant $\widetilde{\Lambda}$.

Indeed, solving Eq.~(\ref{eq:nonconsrho}) yields the standard evolution of the matter energy density,
\begin{equation}
\rho(a) = \rho_{0} \, a^{-3(1 + \omega)},
\end{equation}
which coincides with the behavior in the $\Lambda$CDM model.

Substituting this result into Eq.~(\ref{eq:accel}), one obtains
\begin{equation}
H^2(a) = \frac{\kappa^2}{3} \left( \rho_{0} \, a^{-3(1 + \omega)} + \rho_{\widetilde{\Lambda}} \right),
\end{equation}
which is precisely the Friedmann equation of the standard cosmological model for dark matter and an effective cosmological constant $\widetilde{\Lambda}$.

Therefore, this Ansatz for $Q$ does not introduce genuinely new dynamical features and will not be considered further in the present analysis.

\subsubsection{Power-law Ansatz for the diffusion function}
\label{scale_invariant_Q}

We now consider an extended version of the matter--diffusion scaling solution, in which the matter density is not required to follow the same scaling behavior as the diffusion sector.

For the power-law diffusion model
\begin{equation}
 Q(a) = Q_{0} \; a^{-\mu},
 \label{power_law_Q}
\end{equation}
the continuity equation (\ref{eq:sol_rho_matter}) admits an analytic solution for the matter energy density, yielding
\begin{equation}
\rho \left(a\right) = A ~ a ^{-3(1 + \omega)} + B~ a ^{-\mu},
\label{Eq:rho_m_a}
\end{equation}
with the coefficients $A$ and $B$ defined, respectively, by 
\begin{equation}
A = \rho_{0} ~\left(1 - \frac{\mu Q_{0} / \rho_{0}}{3 (1+\omega) -\mu} \right) \quad \text{and} \quad B = \frac{\mu ~ Q_{0}}{3 (1+ \omega) -\mu},
\label{eq:def_A_B}
\end{equation}
where the positivity of the matter energy density  requires $\mu < 3(1 +\omega ) $, along with with the positive entropy-production condition Eq.~(\ref{eq:good}), which can be recast as:
\begin{equation}
\mu ~ Q_{0} > 0.
\label{thermo_law}
\end{equation}
The solution given by Eq.~(\ref{Eq:rho_m_a}) shows that the matter density is composed of two contributions: the standard dilution term, $\rho \propto a^{-3(1+w)}$, and a diffusion-induced term scaling as $a^{-\mu}$. In general, these two components evolve differently, and the solution describes a superposition of distinct scaling behaviors.

A particularly relevant case arises when $\mu = 3(1+w)$. In this regime, the diffusion function evolves as $Q \propto a^{-3(1+w)}$, and the corresponding solution for the matter energy density, previously discussed in Ref.~\cite{Cruz:2026tjt}, now takes the form
\begin{equation}
\rho(a) = \left[\rho_0 + 3(1+w) Q_0 \ln{a} \right] a^{-3(1+w)}.
\label{limit_ln_case}
\end{equation}
This behavior can be interpreted as a quasi-scaling regime due to the mild logarithmic correction.

Substituting the general solution Eq.~(\ref{Eq:rho_m_a}) into the Friedmann equation and normalizing with respect to the present-day Hubble rate $H_0$, the expansion rate takes the form
\begin{equation}
\frac{H^2(a)}{H_0^2}
=
\Omega_{m,0}\, a^{-3(1+w)}
+
\Omega_{Q,0}\, a^{-\mu}
+
\Omega_{\Lambda_{0}},
\end{equation}
where we have introduced the effective density parameters
\begin{equation}
\Omega_{m,0} \equiv \frac{\kappa^2 A}{3H_0^2},
\quad
\Omega_{Q,0} \equiv
\frac{\kappa^2}{3H_0^2} (B+Q_0),
\quad
\Omega_{\Lambda_{0}} \equiv \frac{\Lambda_{0}}{3H_0^2}.
\end{equation}
This result shows that the cosmological dynamics can be interpreted as an effective three-component system\footnote{Alternatively, the system can be recast as an effective two-component description, consisting of a matter sector subject to energy exchange and a dynamical vacuum component arising from the diffusion contribution. In particular, the scaling $a^{-\mu}$ induces an effective time dependence of the cosmological constant.}. On the other hand, the diffusion sector can be recast as an effective perfect fluid whose energy density scales as $\rho_Q^{\rm eff} \propto a^{-\mu}$. By analogy with the standard scaling $\rho \propto a^{-3(1+w)}$, one may associate to this contribution an effective equation-of-state parameter of the form
\begin{equation}
w^{\rm eff}_{Q} = -1 + \frac{\mu}{3}.
\end{equation}
Thus, the condition for acceleration expansion is simply $\mu < 2$. Depending on the value of $\mu$, the diffusion contribution can mimic different cosmological fluids. In particular, $\mu=0$ corresponds to a cosmological constant with $w^{\rm eff}_{Q}=-1$, $\mu=3$ reproduces a dust-like component with $w^{\rm eff}_{Q}=0$, while $\mu=4$ yields a radiation-like behavior with $w^{\rm eff}_{Q}=1/3$. 
As a general observation, diffusion introduces an additional component with nonstandard scaling, capable of modifying the expansion history and interpolating between different cosmological regimes depending on the underlying diffusion model.

Remarkably, some diffusion functions previously introduced in unimodular gravity from phenomenological or mathematical considerations naturally emerge in the present analysis from the phase-space structure of the autonomous system. This highlights the usefulness of the dynamical systems approach as a systematic framework for restricting admissible diffusion sectors and identifying cosmologically viable diffusion models.

\section{Diffusion reconstruction}\label{sec:dif_rec}

Up to this point, our analysis has focused primarily on the fixed-point structure of the dynamical system, while the physical role of the curvature function $\Gamma_{Q}$ has remained largely implicit. This function ultimately controls the evolution of the diffusion slope $\lambda$ through Eq.~(\ref{eqn:evol_slope_diffusion}), and therefore determines how the diffusion function evolves during the cosmic expansion. In order to make this connection explicit, we now turn to the physical interpretation of $\Gamma_{Q}$ and its role in reconstructing the diffusion function.

Our strategy to reconstruct entire families of diffusion functions $Q(H)$ is similar to potential reconstruction in scalar-field cosmology (see e.g. \cite{Halder:2025ytq}). This is why the diffusion slope parameter resembles the potential slope in scalar-field cosmology.

To explain this idea, we first recall the definition of the diffusion slope by Eq.~(\ref{F_lambda})

\begin{equation}
\frac{d\ln Q}{dN}=-\lambda,
\end{equation}
which allows the equivalent integral representation for $Q(N)$

\begin{equation}
Q(N) = Q_{0} \exp\left[ -\int_{N_{0}}^{N} \lambda(\tilde{N})\, d \tilde{N}\right].
\label{sol_diffusion}
\end{equation}
This expression shows that the diffusion function can be fully reconstructed from the phase-space trajectory through the evolution of the slope parameter $\lambda$. 

An immediate consequence of Eq.~(\ref{sol_diffusion}) is that the sign of $Q$ is fixed by its initial value $Q_{0}$, since the exponential factor is strictly positive. Therefore, within this class of exponential-type solutions, the diffusion function cannot change sign dynamically, which is consistent with the thermodynamics constraint. Any transition between diffusion regimes would require either a vanishing of $Q$ at some stage of the evolution or a more general framework beyond the present formulation.

Alternatively, using the Raychaudhuri equation leads to the master relation
\begin{equation}
\frac{d\ln Q}{d\ln H} = \frac{2\lambda}{3(1+w)\Omega}.
\end{equation}
This equation provides a general differential relation between the diffusion function and the cosmic expansion. Integrating analytically the master equation yields the general reconstruction formula
\begin{equation}
Q(H)=Q_0
\exp \left[ \int_{H_{0}}^{H} \frac{2\lambda(\tilde{H})}{3(1+w)\Omega(\tilde{H})}
\, d\ln \tilde{H} \right].
\label{sol2_diffusion}
\end{equation}
Therefore, Eqs.~(\ref{sol_diffusion}) and (\ref{sol2_diffusion}) provide two equivalent representations for reconstructing diffusion models, either in phase space or in terms of physical variables. In what follows, we will primarily focus on the former description.

So far, we have seen that fixed points of the dynamical system correspond to power-law diffusion solutions when $\lambda$ vanishes or
it is a constant (or equivalently when $\Gamma_{Q}=1$). An important question is whether these local behaviors can be connected by more general diffusion functions for which $\Gamma_Q\neq1$, thereby giving rise to a global cosmological evolution interpolating between different diffusion regimes. 

In what follows, we illustrate the reconstruction framework through a few representative examples, chosen to demonstrate the methodology rather than to provide an exhaustive classification of diffusion models. We emphasize that the formalism is not restricted to the power-law ($\Gamma_Q=1$) or asymptotically constant diffusion models discussed in the literature. Instead, it naturally accommodates a broad class of diffusion functions generated by different choices of the curvature function $\Gamma_Q$. As such, the proposed formalism provides a systematic framework for constructing and classifying diffusion models within unimodular gravity.

\subsubsection{Power-law diffusion solution}

A particularly important situation occurs when $\lambda$ approaches a constant value. This condition corresponds to $\lambda'=0$, which is satisfied nontrivially for the choice $\Gamma_Q=1$. In this case, the evolution equation for $Q$ can be integrated straightforwardly,
\begin{equation}
Q(N)=Q_0\,e^{-\lambda N},
\end{equation}
or equivalently
\begin{equation}
Q(a) = Q_0\, a^{-\lambda}.
\label{power_law_Q_reconstruction}
\end{equation}

Therefore, constant $\lambda$ leads to a power-law diffusion behavior with respect to the scale factor. Physically, this solution corresponds to a regime in which the diffusion sector evolves self-similarly with the cosmological expansion. 

The relation between the diffusion function and the Hubble rate can be obtained using
\begin{equation}
\frac{H'}{H}=-\frac{3}{2}(1+w)\Omega .
\end{equation}
At a fixed point where $\Omega=\Omega_\ast$ is constant, and since we are integrating on an invariant manifold, one obtains
\begin{equation}
H \propto a^{-\frac{3}{2}(1+w)\Omega_\ast} \propto a^{-(1+q_{*})},
\end{equation}
with $q_{*}$ being the deceleration parameter at the fixed point. The above expression exhibits a power-law scaling of the Hubble parameter. This behavior arises naturally within the framework of cosmological dynamical systems, where equilibrium points correspond to asymptotically self-similar solutions. As the system evolves toward a scaling fixed point, the deceleration parameter approaches a constant value, thereby inducing a power-law evolution of the Hubble parameter itself. Consequently, the diffusion sector dynamically acquires a power-law behavior inherited from the asymptotic self-similar structure of the cosmological dynamics \cite{WainwrightEllis1997}.

Moreover, substituting into the power-law solution for $Q$ yields
\begin{equation}
Q(H) \propto H^{\alpha},
\end{equation}
with
\begin{equation}
\alpha = \frac{2\lambda}{3(1+w)\Omega_\ast}.
\end{equation}
In particular, the commonly studied case $Q\propto H^2$ corresponds to the scaling solution where $\lambda = 3(1+w)\Omega_\ast$. In this regime the diffusion sector tracks the matter density and remains a constant fraction of the total energy budget.

The coexistence of the scaling solution with the diffusion-dominated regime is compromised, as anticipated, since the latter requires $\lambda = 0$. Nevertheless, a non-vanishing value today can still be achieved, given the form of Eq.~(\ref{power_law_Q}), before it becomes fully diluted by the cosmic expansion, provided appropriate initial conditions are chosen. This scenario is numerically examined in the next part and in section \ref{sec:dif_rec}.

\subsubsection{Constant-curvature family of diffusion models}

More generally, when the diffusion slope evolves dynamically it does not necessarily follow a simple power-law behavior in the scale factor. This occurs whenever $\Gamma_Q\neq1$, either because it is a constant different from unity or because it depends on the dynamical variables, $\Gamma_Q=\Gamma_Q(\lambda)$. 

We first consider the case in which the curvature parameter is constant, $\Gamma_Q=\Gamma_{0} $, with $\Gamma_{0} \neq 1$. In this situation the phase-space trajectory does not goes sufficiently close to the fixed points identified previously, and the diffusion slope evolves according to Eq.~(\ref{eqn:evol_slope_diffusion}). In this case, integrating the evolution equation yields
\begin{equation}
\lambda(N)=\frac{\lambda_0}{1-(1-\Gamma_{0})\lambda_0 N} ~,
\label{sol_slope}
\end{equation}
which, combined with Eq.~(\ref{sol_diffusion}), leads to the explicit analytical expression
\begin{equation}
Q(N)=Q_0\left[1-(1-\Gamma_{0})\lambda_0 N\right]^{\frac{1}{1-\Gamma_{0}}}.
\label{diffusion_Gamma_const}
\end{equation}

Equation~(\ref{sol_slope}) shows that the diffusion slope does not, in general, correspond to a fixed point of the system, since $\lambda' \neq 0$, except in the special cases $\Gamma_{0}=1$ or $\lambda_0=0$. Instead, constant-curvature models describe trajectories that evolve toward the fixed-point structure of the system but asymptotically approach different cosmological states depending on the value of $\Gamma_{0}$. The behavior of the diffusion function can be classified into three qualitatively distinct regimes according to its curvature:

\begin{itemize}

\item $\Gamma_{0}=1$: Taking the limit $\Gamma_{0}\to 1$ in Eq.~(\ref{diffusion_Gamma_const}) recovers the power-law solution
\begin{equation}
Q(N)=Q_0 e^{-\lambda_{0} N}=Q_0 a^{-\lambda_{0}} ~,
\end{equation}
confirming that the power-law diffusion model arises as a special case of the 
constant-curvature family. As discussed earlier, this solution may approach the scaling matter-diffusion solution when the diffusion slope specifically takes the value $\lambda_{0} = 3(1+w)~\Omega_{*}$.

\item $\Gamma_{0}>1$: In this regime the slope evolution equation $\lambda'=\lambda^2(1-\Gamma_{0})$ implies $\lambda'<0$, so the diffusion slope decreases as the universe expands and asymptotically approaches $\lambda\to0$. Since $Q'=-\lambda Q$, this limit implies $Q'\to 0$ and, therefore, {$Q \to 0$}.  This behavior is also evident directly from Eq.~(\ref{diffusion_Gamma_const}), where the negative exponent causes the dynamical evolution of $Q(N)$ to progressively flatten as $N$ increases, asymptotically approaching zero at late times\footnote{For suitable initial conditions, $Q$ can remain, however, finite but small at the present epoch, while still decaying toward zero in the asymptotic future.}. 

\item $\Gamma_{0} < 1$: In this case $\lambda'>0$, implying that the diffusion slope grows monotonically during the expansion. As a consequence, the diffusion function decays increasingly rapidly with $N$, leading to a strong suppression of the diffusion contribution. However, this suppression does not occur asymptotically. Instead, the solution develops a finite-time singularity, as $\lambda$ diverges at $N_* = [(1-\Gamma_{0})\lambda_0]^{-1}$. Below but near this point, the system is driven toward $Q \to 0$, effectively suppressing diffusion and pushing the dynamics toward the matter--$\Lambda$ sector. Therefore, rather than describing a regular approach to the matter-dominated fixed point, the $\Gamma_{0} < 1$ regime corresponds to a singular trajectory that terminates at the boundary of phase space, signaling a breakdown of the constant-curvature description.

\end{itemize}

Therefore, constant-curvature models do not correspond to genuine fixed points of the full dynamical system and hence do not describe the global phase-space flow associated with the complete cosmological evolution. Instead, they define restricted families of trajectories that asymptotically approach different cosmological regimes depending on the value of $\Gamma_{0}$. 

\begin{figure*}
\centering
\includegraphics[width=0.47\hsize,clip]{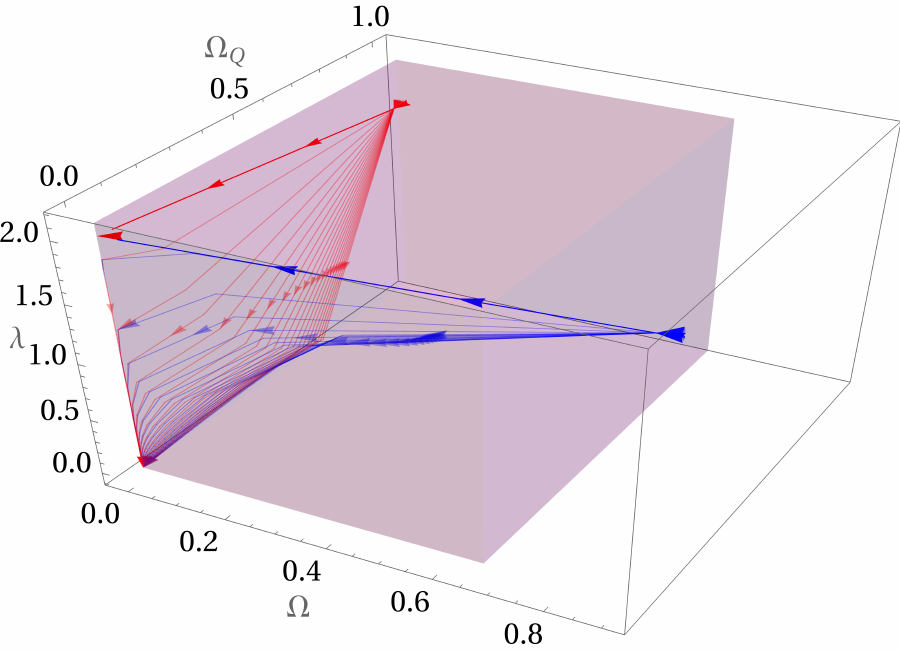}
\includegraphics[width=0.47\hsize,clip]{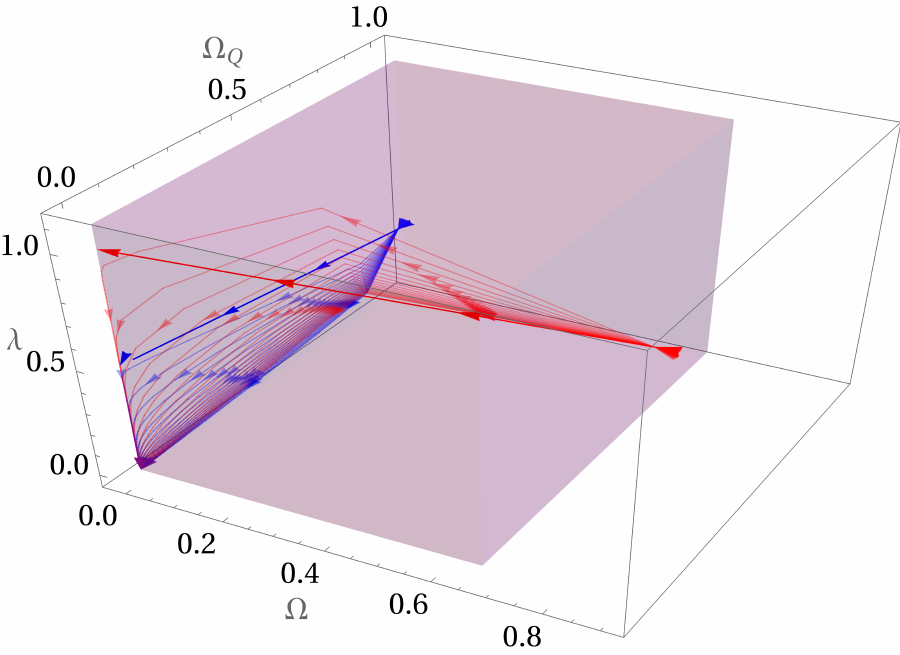}
\caption{Phase-space trajectories for different initial conditions illustrating the dynamical behavior of the constant-curvature family of diffusion models with $\Gamma_0 \geq 1$. Different color tones correspond to distinct values of $\Gamma_0 \in (1,2)$. The case $\Gamma_0=1$ corresponds to constant-$\lambda$ hypersurfaces, effectively reducing the system to a two-dimensional phase space; these trajectories are represented by the darker curves. In the left panel, the trajectories are initialized with $\lambda(t_{\rm ini})=2$, while in the right panel the blue and red trajectories correspond, respectively, to $\lambda(t_{\rm ini})=0.5$ and $\lambda(t_{\rm ini})=1$. The shaded light-purple region indicates the portion of phase space associated with accelerated expansion.} \label{fig:phase_space}
\end{figure*}

To illustrate this behavior, we numerically evolve the system for several initial conditions spanning representative regions of phase space, as shown in Fig.~\ref{fig:phase_space}. In particular, for the range $\Gamma_{0}\in(1,2)$ of interest here, all trajectories exhibit a common qualitative behavior: larger values of $\Gamma_{0}$ lead to stronger deviations from the constant-$\lambda$ regime, driving the system more rapidly toward the asymptotic limit $\lambda\to0$. By contrast, the special case $\Gamma_{0}=1$ corresponds to trajectories evolving on constant-$\lambda$ hypersurfaces, reproducing the power-law diffusion solution discussed previously.

\begin{figure*}
\centering
\includegraphics[width=0.47\hsize,clip]{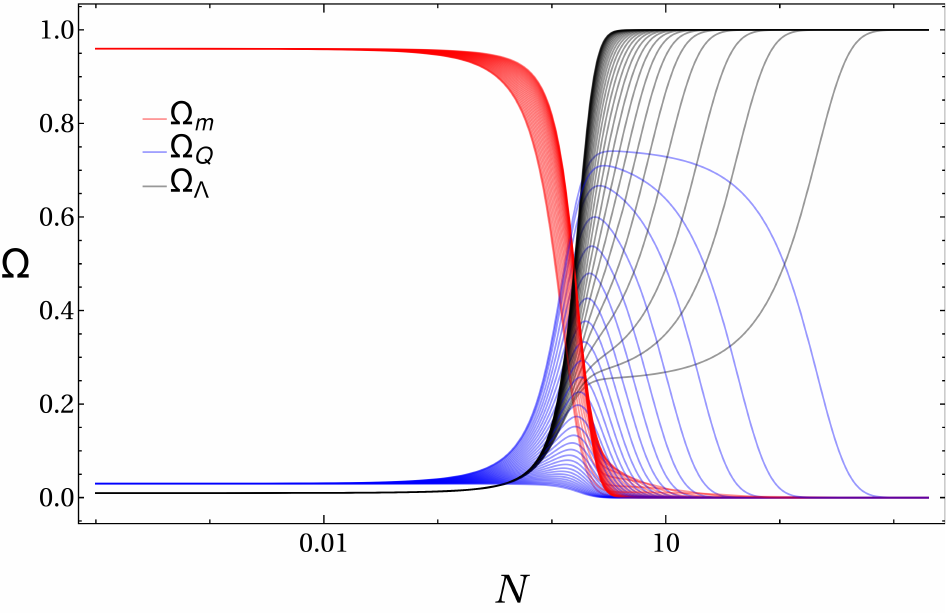}
\includegraphics[width=0.4\hsize,clip]{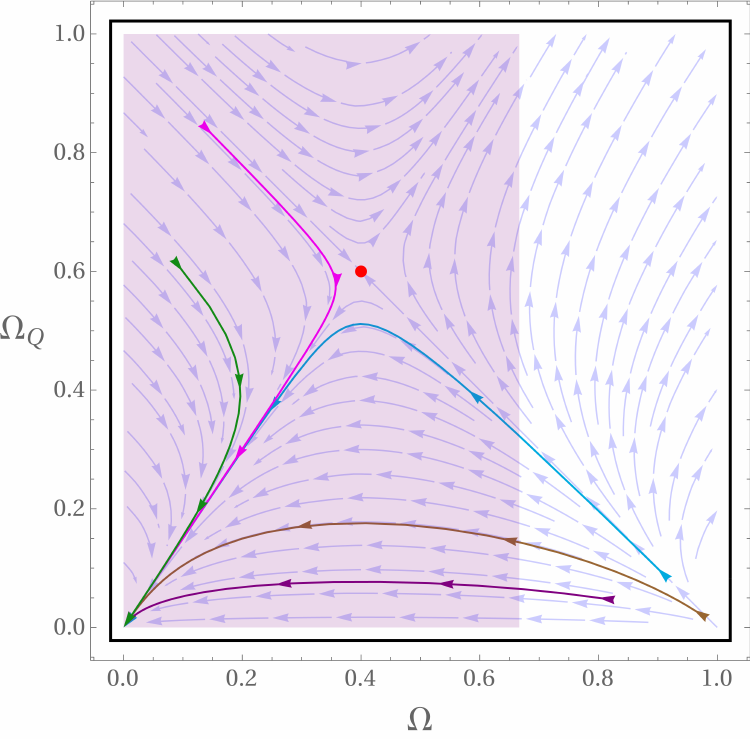}
\caption{Left panel: Numerical evolution of the energy density parameters for the particular case $\Gamma_0=1$ and different values of $\lambda \in (0.01, 3)$. Smaller values of $\lambda$ induce an earlier departure from matter domination and enhance the contribution of the diffusion sector, thereby delaying the onset of the asymptotic $\Lambda$-dominated regime. Right panel: Numerical trajectories in the reduced two-dimensional phase space for the specific case $\lambda=1.2$. The red point denotes the scaling solution located at $\Omega=\lambda/3$ and $\Omega_{Q}=1-\lambda/3$, satisfying the condition $\lambda=3(1+w)\Omega$, with $w=0$. The shaded region indicates the portion of phase space associated with accelerated expansion.} \label{fig:evol_cosmo}
\end{figure*}

To further investigate the $\Gamma_0=1$ regime, left panel of Fig.~\ref{fig:evol_cosmo} displays the evolution of the energy density parameters for different values of $\lambda$, while keeping the initial conditions fixed. Once again, a clear pattern emerges from the numerical evolution: smaller values of $\lambda$ trigger an earlier departure from matter domination and enhance the contribution of the diffusion sector to the accelerated expansion. We have verified that, at the epoch of matter--diffusion equality, the universe already lies within the accelerated-expansion regime, $w_{\rm}<-1/3$. As a consequence, the transition $\Omega \to \Omega_{\Lambda}$ is delayed, with diffusion temporarily playing a significant role before the universe eventually approaches a $\Lambda$-dominated phase at sufficiently late times.

This behavior is further illustrated in the right panel of Fig.~\ref{fig:evol_cosmo}. In particular, the scaling solution (shown as the red point) lies within the accelerated-expansion region, indicating that the universe is already undergoing accelerated expansion at this stage. However, this point does not correspond to an attractor of the system, but rather to a repeller. The true late-time attractor is located at the origin of phase space, $(\Omega,\Omega_Q)=(0,0)$, corresponding to the asymptotic de Sitter solution with $\Omega_{\Lambda}=1$. 

A full dynamical connection between the different cosmological fixed points requires, therefore, a more general curvature function, $\Gamma_Q=\Gamma_Q(\lambda)$, so that the slope evolution equation can vanish at multiple locations in phase space. In this situation the diffusion slope is no longer confined to a single dynamical sector, but can evolve across different regimes during the cosmic expansion. The goal is therefore to construct diffusion functions capable of interpolating between distinct cosmological behaviors, allowing the diffusion term to evolve from a rapidly decaying contribution in the early universe (the scaling solution) to an asymptotically constant component at late times (diffusion dominated).

\subsubsection{General classes of diffusion models}

To illustrate how the diffusion reconstruction framework operates, we consider several representative examples displayed in Table~\ref{tab:diffusion_models}. The central idea is that diffusion models can be generated by specifying the functional form of $\Gamma_Q$, rather than postulating a particular form for the diffusion function $Q(H)$. In this way, the cosmological dynamics of unimodular diffusion models can be systematically classified in terms of the curvature function $\Gamma_Q(\lambda)$. 

The reconstruction procedure follows three steps. First, one specifies a functional form for $\Gamma_Q(\lambda)$. Second, Eq.~(\ref{eqn:evol_slope_diffusion}) is integrated to determine the evolution of the diffusion slope $\lambda(N)$. Finally, this result is substituted into Eq.~(\ref{sol_diffusion}) to reconstruct the corresponding diffusion function.

In practice, the slope equation must admit multiple non-trivial zeros. The simplest strategy is therefore to construct a function $\Gamma_{Q}=\Gamma_Q(\lambda)$ that crosses unity at least once, thereby ensuring the existence of non-trivial stationary configurations. In this way, the system can dynamically interpolate between the matter-dominated, scaling, and diffusion-dominated regimes. In the following, we explore some representative choices and their physical implications:

\begin{itemize}

\item A simple example is the linear model
\begin{equation}
\Gamma_Q(\lambda)=1+\alpha(\lambda-\lambda_{s}),
\end{equation}
where $\alpha$ and $\lambda_{s}$ are constants. The invertibility condition on $\lambda(N)$ requires the evolution to remain on a single branch, either $\lambda>\lambda_s$ or $\lambda<\lambda_s$, without crossing the stationary point $\lambda=\lambda_s$. According to Eq.~(\ref{eqn:evol_slope_diffusion}), the values $\lambda=0$ and $\lambda=\lambda_s$ correspond to stationary solutions of the slope dynamics. The former yields a constant diffusion function, $Q=Q_0$, associated with the diffusion-dominated regime, whereas the latter reproduces the power-law behavior $Q\propto a^{-\lambda_s}$ characteristic of the scaling solution. Although $\lambda_s$ does not affect the location of the fixed points of the autonomous system, it controls the evolution of the diffusion slope and therefore the way cosmological trajectories approach the asymptotic regimes.

The reconstructed diffusion function, valid away from the stationary solution
$\lambda=\lambda_s$, can be expressed parametrically as
\begin{equation}
Q(\lambda)
=
Q_{0}
\left(
1-\frac{\lambda_{s}}{\lambda}
\right)^{\frac{1}{\alpha \lambda_{s}}},
\label{Eq:Q_lambda_linear}
\end{equation}
which is ill-defined for $\lambda = 0$. However, this corresponds to the degenerate case satisfying Eq.~(\ref{eqn:evol_slope_diffusion}), namely a fixed point of the evolution equation ${\lambda}^{\prime}=0$. Otherwise, it describes a generalized power-law diffusion sector.
\item For the inverse model 
\begin{equation}
\Gamma_Q(\lambda)=1+\frac{\alpha}{\lambda},
\end{equation}
the diffusion slope  evolves as $\lambda=\lambda_0 e^{-\alpha N}$, leading to qualitatively different behaviors depending on the sign of $\alpha$. The corresponding diffusion function is 
\begin{equation}
Q(N)
=
Q_{0}
\exp\!\left[
-\frac{\lambda_{0}}{\alpha}
\left(1-e^{-\alpha N}\right)
\right].
\label{Eq: Q_exponential_reconstruction}
\end{equation}
For $\alpha>0$, the slope decays exponentially and the diffusion function asymptotically approaches a finite constant, $Q \to Q_0 \exp(-\lambda_0/\alpha)$, effectively freezing the diffusion sector. Nevertheless, note that $\Gamma_Q$ itself diverges in the large-$N$ limit. In contrast, for $\alpha<0$, the slope grows exponentially, driving the diffusion function to zero in a super-exponential fashion. This results in a transient diffusion-dominated phase followed by a rapid suppression of diffusion, after which the cosmological evolution is driven toward a de Sitter phase dominated by the cosmological constant. The magnitude of $\alpha$ controls the efficiency of this suppression, with smaller $|\alpha|$ leading to a more pronounced intermediate regime.

In this case ($\lambda'=-\alpha \lambda$), $\lambda=0$ is the only finite fixed point, while the condition $\Gamma_Q=1$ is satisfied only asymptotically as $\lambda \to \infty$. Consequently, the standard scaling solution is not realized, and the dynamics is effectively restricted to the invariant-manifold fixed points associated with $\lambda=0$.

\end{itemize}

\begin{table*}[htp]
\centering
\begin{tabular}{c c c c}
\hline
$\Gamma_Q(\lambda)$ & Diffusion slope $\lambda(N)$ & Diffusion law $Q(N)$ & Physical behavior \\
\hline

$1$ &
$\lambda=\mu ~(\mathrm{const})$ &
$Q=Q_0 a^{-\mu}$ &
Power-law diffusion \\

$\gamma$ (const) &
$\lambda=\dfrac{\lambda_0}{1-(1-\gamma)\lambda_0 N}$ &
$Q=Q_{0} [1-(1-\gamma)\lambda_{0}N]^{\frac{1}{1-\gamma}}$ &
Running diffusion slope \\

$1+\alpha(\lambda-\lambda_{s})$ &
$(1-\frac{\lambda_{s}}{\lambda})^{\frac{1}{\lambda_s^2}} \exp{\frac{1}{\lambda_s \lambda} }= C a^{-\alpha}$ &
$Q(\lambda)
=
Q_{0}
\left(
1-\frac{\lambda_{s}}{\lambda}
\right)^{\frac{1}{\alpha \lambda_{s}}}$ &
Generalized power-law diffusion \\

$1+\alpha/\lambda$ &
$\lambda=\lambda_0 e^{-\alpha N}$ &
$Q=Q_0\exp\!\left[-\frac{\lambda_0}{\alpha}(1-e^{-\alpha N})\right]$ &
Asymptotically constant diffusion \\

\hline
\end{tabular}
\caption{Classes of diffusion models generated by different choices of the curvature function $\Gamma_Q$. Each choice determines the evolution of the diffusion slope $\lambda$ and therefore reconstructs the diffusion function $Q$.}\label{tab:diffusion_models}
\end{table*}

In this way, one can systematically construct diffusion functions that incorporate multiple dynamical regimes within a single cosmological evolution while respecting the mathematical constraints of the dynamical system. The resulting framework provides a promising setting for exploring the cosmological implications of diffusion processes and for identifying viable models to be confronted with observational data.

\section{Conclusions}
\label{sec:conclusions}

A central aspect of the present work was the identification of the invertibility condition that the diffusion slope must satisfy in order to generate a closed autonomous cosmological system. Although a physical diffusion sector is not required, in general, to satisfy this condition, it nevertheless provides a robust and systematic strategy for constructing diffusion models from a dynamical system perspective.  We therefore focused on this broad class of diffusion functions, which remains sufficiently general to describe a rich cosmological dynamics. 

Under these conditions, we determined the physically admissible fixed points governing the cosmic expansion. In particular, we identified a matter--diffusion scaling regime associated with power-law diffusion, together with a purely diffusion-dominated solution capable of driving late-time accelerated expansion without requiring an explicit cosmological constant term.

Remarkably, the present framework naturally reproduces diffusion models previously introduced in the UG literature from phenomenological considerations, including barotropic and power-law diffusion models. The present analysis shows that these diffusion functions emerge directly from the phase-space structure of the dynamical system.

After establishing the local phase-space dynamics, we investigated the cosmological implications of extending the diffusion behaviors obtained near the fixed points to the entire cosmological evolution. This provides a well-defined framework for exploring specific diffusion scenarios characterized by distinct global behaviors.

In the absence of a fundamental principle selecting the physically preferred diffusion function, we also developed a reconstruction framework in which viable diffusion sectors are determined directly from the cosmological dynamics. Building upon the phase-space structure of the system, this formalism allows for cosmological trajectories beyond the asymptotic fixed-point regimes, including models capable of interpolating between different diffusion behaviors during the cosmic evolution. In this way, the reconstruction program provides a systematic procedure for generating broad classes of diffusion cosmologies (see Table \ref{tab:diffusion_models}) while preserving the autonomous structure of the dynamical system. This analysis was further complemented by a numerical exploration of the phase space, illustrating the global behavior of the cosmological trajectories and their associated dynamical sectors.

The formalism developed here also sets the basis for a detailed observational analysis aimed at constraining the admissible diffusion sectors and selecting physically viable cosmological scenarios. Such an observational scrutiny constitutes the natural next step of the present program.

Beyond the background evolution, the diffusion mechanism is also expected to produce distinctive signatures at the perturbative level. Since diffusion modifies the continuity equations through energy exchange within the dark sector, it can affect the evolution of matter density perturbations and consequently the growth of cosmic structures. These effects may leave observable imprints on the matter power spectrum and the growth rate of structures, providing a potential avenue to distinguish diffusion cosmologies from the standard $\Lambda$CDM scenario. Their magnitude and sign depend on the specific diffusion law and the efficiency of the energy transfer. A detailed perturbative analysis is left for future work.

\section*{Acknowledgments}
Financial support from the Chilean National Agency for Research and Development (ANID) through Fondecyt Grant No. 1250969 is gratefully acknowledged.

\bibliographystyle{ieeetr}
\bibliography{biblio.bib}
\end{document}